\documentclass[twocolumn,showpacs,prb,amsmath,amssymb]{revtex4}
\usepackage{graphicx}% Include figure files
\usepackage{dcolumn}% Align table columns on decimal point
\usepackage{bm}% bold math
\begin{document}
\title{Impact of the electron-electron correlation on phonon dispersions:\\
failure of LDA and GGA functionals in graphene and graphite}
\author{Michele Lazzeri$^1$, Claudio Attaccalite$^2$, Ludger Wirtz$^3$,  and Francesco Mauri$^1$}
\affiliation{$^1$ IMPMC, Universit\'es Paris 6 et 7, CNRS, IPGP, 140 rue de Lourmel, 75015 Paris, France }
\affiliation{$^2$ ETSF European Theoretical Spectroscopy Facility and  Universidad del Pais Vasco, Unidad de Fisica de Materiales, San Sebastian, Spain }
\affiliation{
$^3$ Institute for Electronics, Microelectronics, and Nanotechnology, CNRS, 59652 Villeneuve d'Ascq, France
}
\date{\today}

\begin{abstract}
We compute electron-phonon coupling (EPC) of selected phonon modes
in graphene and graphite using various ab-initio methods.
The inclusion of non-local exchange-correlation effects within the
GW approach strongly renormalizes the
square EPC of the A$'_1$ {\bf K} mode by almost 80$\%$
with respect to density functional theory in the LDA and GGA
approximations.
Within GW, the phonon slope of the A$'_1$ {\bf K} mode is almost two times
larger than in GGA and LDA, in agreement with phonon dispersions
from inelastic x-ray scattering and Raman spectroscopy.
The hybrid B3LYP functional overestimates the EPC at {\bf K} by about 30\%.
Within the Hartree-Fock approximation, the graphene structure 
displays an instability under a distortion following the A$'_1$ phonon 
at {\bf K}.
\end{abstract}

%%      71.15.Mb   Density functional theory, local density approximation,
%%                 gradient and other corrections
%%      63.20.kd   Phonon-electron interaction
%%      78.30.Na   Infrared and Raman spectra: Fullerenes and related materials
%%      81.05.Uw   Carbon, diamond, graphite

\maketitle
The electron-phonon coupling (EPC) is one of the fundamental
quantities in condensed matter. It determines phonon-dispersions and
Kohn anomalies, phonon-mediated superconductivity, electrical
resistivity, Jahn-Teller distortions etc.  Nowadays, density
functional theory within local and semi-local approximations
(DFT) is considered the ''standard model'' to
compute ab-initio the electron-phonon interaction and phonon
dispersions~\cite{DFPT0}.  Thus, a failure of DFT would have major
consequences in a broad context.   
In GGA and LDA approximations~\cite{note_defs}, the
electron exchange-correlation energy is a local functional of the
charge density and the long-range character of the electron-electron
interaction is neglected.
These effects are taken into account by
Green-function approaches based on the screened electron-electron
interaction W, such as the GW method~\cite{GWreview}.  GW is
considered the most precise ab-initio approach to determine electronic
bands but, so far, it has never been used to compute EPCs nor phonon
dispersions.
The semi-empirical B3LYP functional~\cite{note_defs} partially includes 
long-range
Hartree-Fock exchange. B3LYP has been used to compute phonon frequencies
but, so far, not the electron-phonon coupling.

The electron-phonon coupling is a key quantity for graphene, graphite
and carbon nanotubes.  It determines the Raman spectrum, which is the
most common characterization technique for graphene and
nanotubes~\cite{reich04,ferrari06}, and the high-bias electron
transport in nanotubes~\cite{yao00}.  Graphene and graphite are quite
unique systems in which the actual value of the EPC for some phonons
can be obtained almost directly from measurements.  In particular, the
square of the EPC of the highest optical phonon branch (HOB) at the
symmetry {\bf K}-point is proportional to the HOB slope near {\bf
K}~\cite{piscanec04}.  The HOB {\bf K } slope can be measured by
inelastic x-ray scattering (IXS)~\cite{maultzschPRL04,mohr07} or by
the dispersion of the D and 2D lines as a function of the excitation
energy in a Raman experiment
~\cite{pocsik98,tan02,maultzschPRB04,thomsen00,ferrari06}.  A careful
look at the most recent data suggests that the experimental phonon
slopes (and thus the EPC) are underestimated by DFT~\cite{ferrari06}.
The ability of DFT (LDA and GGA)
in describing the EPC of graphene was also
questioned by a recent theoretical work~\cite{basko08}.

Here, we show that: i) the GW approach, which provides
the most accurate ab-initio treatment of electron-correlation,
can be used to compute the electron-phonon interaction and the phonon
dispersion;
ii) in graphite and graphene, DFT (LDA and GGA)
underestimates by a factor 2 the slope of the phonon dispersion of
the highest optical branch at the zone-boundary and the
square of its electron-phonon coupling by almost 80\%;
iii) GW reproduces both the experimental phonon dispersion near {\bf K}, the
value of the EPC and the electronic band dispersion;
iv) the B3LYP hybrid functional~\cite{note_defs} gives phonons close to GW
but overestimates the EPC at {\bf K} by about 30 \%;
v) within Hartree-Fock the graphite structure is unstable.

\begin{figure} [h]
\centerline{\includegraphics[width=65mm]{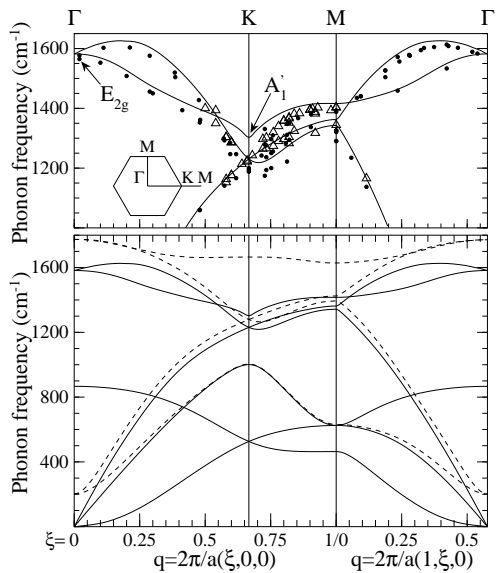}}
\caption{Upper panel: Phonon dispersion of graphite.
Lines are DFT calculations,
dots and triangles are IXS measurements
from Refs.~\onlinecite{maultzschPRL04,mohr07}, respectively.
Lower panel: phonon dispersion of graphene from DFT calculations.
Dashed lines are 
obtained by subtracting from the dynamical matrix the
phonon self-energy between the $\pi$ bands ($\widetilde{\omega_{\bf q}}$
in the text).}
\label{fig1}
\end{figure}

In Fig.~\ref{fig1}, we show the phonon dispersion of graphite computed
with DFT$_{\rm GGA}$ ~\cite{DFPT1}.  In spite of the general good agreement with IXS
data, the situation is not clear for the HOB near {\bf K}.  In
fact, despite the scattering among experimental data, the theoretical
HOB is always higher in energy with respect to measurements and the
theoretical phonon slope (for the HOB near {\bf K}) is underestimating
the measured one.  It is also remarkable that while the DFT
{\bf K} frequency is $\sim$~1300 cm$^{-1}$, the highest
measured is much lower at $\sim$~1200 cm$^{-1}$.

\begin{figure} [h]
\centerline{\includegraphics[width=55mm]{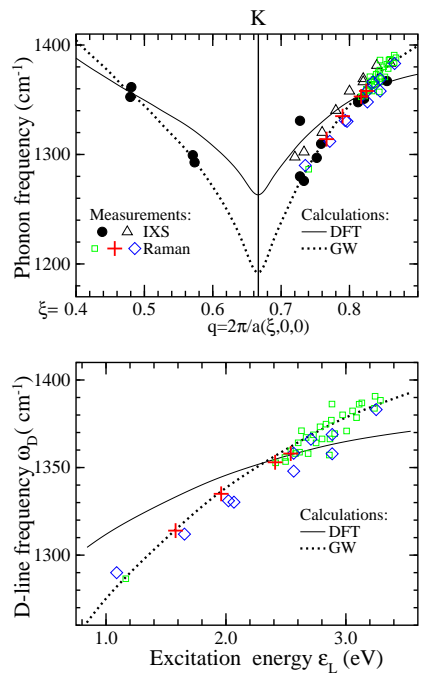}}
\caption{(Color online).
Upper panel: dispersion of the highest optical phonon in graphite near {\bf K}.
Calculations are from DFT, or corrected to include GW
renormalization of the electron-phonon coupling.
Here, the DFT dispersion is vertically shifted by 
-40~cm$^{-1}$ to fit measurements.
Dots and triangles are IXS data
from Refs.~\onlinecite{maultzschPRL04,mohr07}, respectively.
Squares, plus and diamonds are obtained 
from Raman data of Refs.
~\onlinecite{pocsik98,tan02,maultzschPRB04}, respectively,
using the double resonance model~\cite{thomsen00,maultzschPRB04}.
Lower panel: dispersion of the Raman D-line.}
\label{fig2}
\end{figure}

The dispersion of the HOB near {\bf K} can also be obtained by Raman
measurements of the graphene and graphite D-line
($\sim$1350~cm$^{-1}$)~\cite{maultzschPRB04}.
The D-line frequency $\omega_D$ depends on the energy of the exciting
laser $\epsilon_L$.
According to the double-resonance model~\cite{thomsen00,maultzschPRB04},
$\epsilon_L$ activates a phonon of the
HOB with momentum {\bf q}={\bf K}+$\Delta${\bf q} along the
{\bf K}-{\bf M} line~\cite{ferrari06}
and energy $\hbar\omega_D$. $\Delta${\bf q} is determined by
$\epsilon_{{\bf K}-\Delta{\bf q},\pi^*}-
\epsilon_{{\bf K}-\Delta{\bf q},\pi} = \epsilon_L - \hbar\omega_D/2,$
where $\epsilon_{{\bf k},\pi/\pi^*}$ is the energy of the $\pi/\pi^*$
electronic state with momentum {\bf k}.
Thus, by measuring $\omega_D$ vs. $\epsilon_L$ and considering the 
electronic $\pi$ bands dispersion from DFT one can obtain the phonon
dispersion $\omega_D$ vs. {\bf q}~\cite{maultzschPRB04}.
The phonon dispersion thus obtained is very similar to the one from
IXS data and its slope is clearly underestimated by DFT (Fig.~\ref{fig2}, upper panel). 
The same conclusion is reached by comparing the D-line dispersion
$\omega_D$ vs. $\epsilon_L$ (directly obtained from measurements)
with calculations (Fig.~\ref{fig2}, lower panel).
Note that the dispersions of the Raman 2D-line~\cite{ferrari06} is consistent
with the dispersion of the D and thus in disagreement with DFT (LDA and GGA)
as well.

The steep slope of the HOB near {\bf K} is due to the presence of a
Kohn anomaly for this phonon~\cite{piscanec04}.
In particular, in Ref.~\onlinecite{piscanec04}, it was shown that the
HOB slope is entirely determined by the contribution of the phonon
self-energy between $\pi$-bands, $P_{\bf q}$,
to the  dynamical matrix, $\cal{D}_{\bf q}$.
$\omega_{\bf q}=\sqrt{{\cal D}_{\bf q}/m}$ is the phonon pulsation, where
$m$ is the mass.
For a given phonon with momentum {\bf q},
\begin{equation}
{\cal D}_{\bf q}=B_{\bf q}+P_{\bf q}~~;~~
P_{\bf q} = \frac{4}{N_k}\sum_{{\bf k}}
\frac{|D_{({\bf k+q})\pi^*,{\bf k}\pi}|^2}
{\epsilon_{{\bf k},\pi}-\epsilon_{{\bf k+q},\pi^*}}
\label{eq1}
\end{equation}
where 
the sum is performed on $N_k$ wavevectors all over the Brillouin zone,
$D_{({\bf k+q})i,{\bf k}j} = 
\langle {\bf k+q},i|\Delta V_{\bf q}|{\bf k},j\rangle$
is the EPC, $\Delta V_{\bf q}$ is the derivative of the Kohn-Sham
potential with respect to the phonon mode,
$|{\bf k},i\rangle$ is the Bloch eigenstate with momentum {\bf k},
band index $i$ and energy $\epsilon_{{\bf k},i}$.
$\pi$($\pi^*$) identifies the occupied (empty) $\pi$-band.
In Fig.~\ref{fig1} we show a fictitious phonon
dispersion $\widetilde{\omega_{\bf q}}$ obtained subtracting $P_{\bf q}$ from the dynamical matrix
($\widetilde{\omega_{\bf q}}=\sqrt{B_{\bf q}/m}$) for each phonon.
The HOB is the branch which is mostly affected and, for the HOB,
$\widetilde{\omega_{\bf q}}$ becomes almost flat near {\bf K}.
Thus, DFT (LDA or GGA) fails in describing the HOB slope near {\bf K}, slope which
is determined by $P_{\bf q}$. $P_{\bf q}$ is given by the the square
EPC divided by $\pi$-band energies.
Thus, the DFT failure can be attributed to a poor description of the EPC
or of the $\pi$-band dispersion.

In graphene and graphite, it is known that standard DFT provides an
underestimation of the $\pi$ and $\pi^*$-band slopes of
$\sim10-20$\%~\cite{zhou05,gruneis08}.
A very precise description of the bands, in better agreement with measurements,
is obtained using GW~\cite{zhou05,gruneis08}.
We thus computed the $\pi$-bands with DFT (both LDA and GGA)~\cite{PBE_LDA}
and GW~\cite{GW} and compared with 
Hartree-Fock (HF)~\cite{CRYSTAL} and B3LYP~\cite{CRYSTAL}. Details are
in ~\onlinecite{comp_det}.  The different methods provide band
dispersions whose overall behavior can be described by a scaling of
the $\pi$ energies~\cite{zhou05}.  The different scaling factors can
be obtained by comparing $\Delta\epsilon_g$: the energy difference
between the $\pi^*$ and $\pi$ bands at the symmetry point {\bf M} ({\bf
L}) for graphene (graphite).  $\Delta\epsilon_g$ is larger in
GW than in DFT (Tab.~\ref{tab1}).  Thus, inclusion of the
GW correction to the electronic bands alone results in a larger denominator
in Eq.~\ref{eq1}, providing a smaller phonon slope and a worse
agreement with experiments.  The underestimation of the {\bf K}
phonon slope in DFT is, thus, due to the EPC.

\begin{figure} [h]
\centerline{\includegraphics[width=60mm]{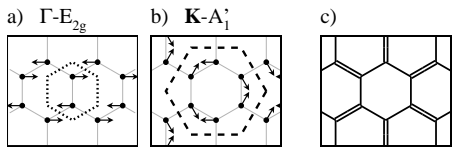}}
\caption{a, b): patterns of the ${\bm \Gamma}$-E$_{2g}$ and
{\bf K}-A$'_1$ phonons of graphene. Dotted and dashed lines are the
Wigner-Seitz cells of the unit-cell and of the $\sqrt{3}\times\sqrt{3}$
super-cell. c): Hartree-Fock equilibrium structure.
}
\label{fig3}
\end{figure}

The EPC can be computed with linear response as, e.g., in
Ref.~\onlinecite{piscanec04} but, at present, the use of this
technique within GW is not feasible.  Alternatively, the EPC
associated to a phonon mode can be determined by the variation of the
electronic band energies by displacing the atoms according to the
considered mode.  In graphene, at {\bf K}, there are doubly degenerate
$\pi$ electronic states at the Fermi level.  The HOB
corresponds to the E$_{2g}$ phonon at ${\bm \Gamma}$ and to the A$'_1$
at {\bf K}.  As an example, we consider the EPC associated to the
${\bm \Gamma}$-E$_{2g}$ phonon and we displace the atoms according to
its phonon pattern (see Fig.~\ref{fig3}). Following symmetry
arguments~\cite{slonczewski58}, one can show that, in an arbitrary
base of the two-dimensional space of the $\pi$ bands at {\bf K}, the
Hamiltonian is the 2$\times$2 matrix:
\begin{equation}
H = 2\sqrt{\langle D^2_{\bm\Gamma}\rangle_{\rm F}}
\begin{pmatrix} a&b\\b^*&-a\end{pmatrix}d+\mathcal{O}(d^2),
\label{eq2}
\end{equation}
where {\it each atom is displaced by}~$d$, $|a|^2+|b|^2=1$, and
$\langle D^2_{\bm
\Gamma}\rangle_{\rm F} = \sum_{i,j}^{\pi,\pi^*} |D_{{\rm \bf K}i,{\rm \bf
K}j}|^2/4$, where the sum
is performed on the two degenerate $\pi$ bands.
Diagonalizing Eq.~\ref{eq2}, we see that an atomic displacement
following the ${\bm \Gamma}$-E$_{2g}$
phonon induces the splitting
$\Delta E_{\bm \Gamma}=\epsilon_{{\bf K},{\pi^*}}-
\epsilon_{{\bf K},{\pi}}$ and 
\begin{equation}
\langle D^2_{\bm \Gamma}\rangle_{\rm F} = \lim_{d\rightarrow 0}
\frac{1}{16}
\left(\frac{\Delta E_{\bm \Gamma}}{d}\right)^2.
\label{eqEPC1}
\end{equation}
In analogous way, we define 
$\langle D^2_{\bf K}\rangle_{\rm F} = \sum_{i,j}^{\pi,\pi^*} |D_{(2{\rm
\bf K})i,{\rm \bf K}j}|^2/4$ for the A$'_1$ phonon at {\bf K}.
Let us consider a $\sqrt{3}\times\sqrt{3}$ graphene supercell.
Such a cell can be used to displace the atoms following the
{\bf K}-A$'_1$ phonon (Fig.~\ref{fig3}), since
the {\bf K} point is refolded in ${\bm \Gamma}$.
Let us call $\Delta E_{\bf K}$ the splitting of the $\epsilon_{{\bf K},{\pi}}$
bands induced by this displacement
(since {\bf K} is refolded in ${\bm \Gamma}$, here $\epsilon_{{\bf K},\pi}$
denotes the energies of the ${\bm \Gamma}$ band of the supercell
corresponding to the $\pi$ band at {\bf K} in the unit cell).
Considering the atomic distortion of Fig.~\ref{fig3},
displacing each atom by $d$, one can show that
\begin{equation}
\langle D^2_{\bf K}\rangle_{\rm F} = \lim_{d\rightarrow 0} \frac{1}{8}
\left(\frac{\Delta E_{\bf K}}{d}\right)^2.
\label{eqEPC2}
\end{equation}
In practice, by calculating band energies in the distorted
structures of Fig.~\ref{fig3} and using Eqs.~\ref{eqEPC1},~\ref{eqEPC2}
one obtains the EPCs of the ${\bm \Gamma}$-E$_{2g}$ and
{\bf K}-A$'_1$ phonons between $\pi$ states.
Similar equations can be used for graphite~\cite{note01}.
Results are in Tab.~\ref{tab1} together with the computed phonon frequencies.
The EPCs from DFT$_{\rm GGA}$
are in agreement with those from linear response~\cite{piscanec04}.
We also remark that, within the present "frozen-phonon" approach, the
Coulomb vertex-corrections are implicitly included within GW.

To study the effect of the different computational methods on the
the phonon slope (which is determined by $P_{\bf q}$) we recall that
$P_{\bf q}$ is the ratio of the square EPC and band energies (Eq.~\ref{eq1}).
Thus, we have to compare 
$\alpha_{\bf q}=\langle D^2_{\bf q}\rangle_{\rm F} / \Delta\epsilon_g$.
As an example, assuming that the change of $P_{\bf q}$ from DFT to GW
is constant for {\bf q} near {\bf K},
\begin{equation}
\frac{P^{GW}_{\bf q}}{P^{DFT}_{\bf q}}\simeq
\frac{\alpha^{GW}_{\bf K}}{\alpha^{DFT}_{\bf K}}=r^{GW}
\label{eq5}
\end{equation}
and $r^{GW}$ provides the change in the {\bf K} phonon slope going
from DFT to GW.  
To understand the results, we recall that in standard DFT the 
exchange-correlation depends only on the local electron-density.
In contrast, the exchange-interaction in HF and GW is non-local.
Furthermore, in GW, correlation effects are 
non-local since they are described through a dynamically screened
Coulomb interaction. The hybrid functional B3LYP gives results
intermediate between DFT and HF.

\begin{table}
\centering
\caption{Electron-phonon coupling of the ${\bm \Gamma}$-E$_{2g}$ and
{\bf K}-A$'_1$ phonons computed with various approximations.
$\Delta\epsilon_g$ (eV),
$\langle D^2_{\bf q}\rangle_{\rm F}$ (eV$^2$/\AA$^2$)
and $\alpha_{\bf q}$ (eV/\AA$^2$)
are defined in the text.
$\omega_{\bm \Gamma}$ ($\omega_{\bf K}$) is the phonon frequency
of the E$_{2g}$ (A$'_1$) mode (cm$^{-1}$).
The GW $\omega_{\bf K}$ for graphite (in parenthesis)
is not computed directly (see the text).
$i=\sqrt{-1}$ is the imaginary unit.
}
\label{tab1}
\begin{tabular}{l|ccccccc}
    &\multicolumn{7}{c}{Graphene:}\\
    & $\Delta\epsilon_g$                        &
      $\langle D^2_{\bm \Gamma}\rangle_{\rm F}$ &
      $\alpha_{\bm \Gamma}$                     &
      $\omega_{\bm \Gamma}$                     &
      $\langle D^2_{\bf K}\rangle_{\rm F}$      &
      $\alpha_{\bf K}$                          &
      $\omega_{\bf K}$                          \\
DFT$_{\rm LDA}$ & 4.03  &  44.4   & 11.0 & 1568   &   89.9  & 22.3 &   1275          \\
DFT$_{\rm GGA}$ & 4.08  &  45.4   & 11.1 & 1583   &   92.0  & 22.5 &   1303          \\
GW              & 4.89  &  62.8   & 12.8 &   --   &  193    & 39.5 &     --          \\
B3LYP           & 6.14  &  82.3   & 13.4 & 1588   &  256    & 41.7 &  1172           \\
HF              & 12.1  &  321    & 26.6 & 1705   & 6020    & 498  &  960$\times i$  \\
\hline
    &\multicolumn{7}{c}{Graphite:}                          \\
    & $\Delta\epsilon_g$                                    &
      $\overline{\langle D^2_{\bm \Gamma}\rangle}_{\rm F}$  &
      $\alpha_{\bm \Gamma}$                                 &
      $\omega_{\bm \Gamma}$                                 &
      $\overline{\langle D^2_{\bf K}\rangle}_{\rm F}$       &
      $\alpha_{\bf K}$                                      &
      $\omega_{\bf K}$                                      \\
DFT$_{\rm LDA}$ & 4.06  &  43.6   & 10.7 &  1568  &   88.9  & 21.8 &   1299          \\
DFT$_{\rm GGA}$ & 4.07  &  44.9   & 11.0 &  1581  &   91.5  & 22.5 &   1319          \\
GW              & 4.57  &  58.6   & 12.8 &    --  &  164.2  & 35.9 &  (1192)         \\
\end{tabular}
\end{table}

Both $\alpha_{\bm \Gamma}$ and $\alpha_{\bf K}$ are
heavily overestimated by HF, the {\bf K}-EPC being so huge that
graphene is no more stable (the {\bf K}A$'_1$ phonon frequency is not
real).  Indeed, the HF equilibrium geometry 
is a $\sqrt{3}\times\sqrt{3}$ reconstruction with alternating double and single
bonds of 1.40 and 1.43~\AA~lengths as in Fig.~\ref{fig3}
(with a gain of 0.9 meV/atom).
These results demonstrate the major effect of the
long-range character of the exchange for the {\bf
K}-EPC~\cite{basko08} but also the importance of the proper inclusion
of the screening (included in GW but neglected in HF).
Notice also that $\alpha^{GW}_{\bf K}$ of graphite is smaller with respect
to graphene by $\sim$10\%.  This is explained by the larger
screening of the exchange in graphite 
(due to the presence of adjacent layers) than in graphene.
On the contrary, within GGA and LDA, the graphite phonon frequencies
and EPCs are very similar to those of graphene, since these functionals do 
not take into account the electron-electron interaction screening.
%Given the major importance of the electronic screening, precise and 
%parameter-free calculations can rely only on a fully ab-initio treatment 
%such as GW and not on model Hamiltonians.

Concerning the phonon slope, $\alpha_{\bm \Gamma}^{GW}$ is 15\% larger than
$\alpha_{\bm \Gamma}^{DFT}$. Indeed, DFT reproduces with
this precision the phonon frequency and dispersion of the
HOB at ${\bm \Gamma}$. On the contrary, $\alpha^{GW}_{\bf K}$ is 60\% larger
than $\alpha^{DFT}_{\bf K}$, for graphite.
This large increase with respect to DFT
could explain the disagreement between DFT and the
measured  A$'_1$ phonon dispersion near {\bf K}.
To test this, we need to determine the GW phonon dispersion
that, using Eq.~\ref{eq5} becomes
$\omega^{GW}_{\bf q}\simeq\sqrt{(B^{GW}_{\bf q}+r^{GW}P^{DFT}_{\bf q})/m}$,
where $r^{GW}=1.6$.
Moreover, we can assume $B^{GW}_{\bf q}\simeq B^{GW}_{\bf K}$ since
the $B_{\bf q}$ component of the
dynamical matrix (Eq.~\ref{eq1}) is not expected to have an important
dependence on {\bf q} (Fig.~\ref{fig1}).
The value of $B^{GW}_{\bf K}$ is obtained 
as a fit to the measurements of Fig.~\ref{fig2}~\cite{cumber}. 
The resulting {\bf K} A$'_1$ phonon frequency 
is 1192~cm$^{-1}$ which is our best estimation
and is almost 100~cm$^{-1}$ smaller than in DFT.
The  phonon dispersion thus obtained and the corresponding D-line dispersion
are both in better agreement with measurements (Fig.~\ref{fig2}).
%We remark that this procedure is expected to provide a good
%description of the HOB phonon slope near {\bf K}, since the dispersion
%is almost entirely determined by the phonon self-energy $P_{\bf q}$
%(Fig.~\ref{fig1} and Ref.~\onlinecite{piscanec04}).

The partial inclusion of long-range exchange
within the semiempirical B3LYP functional leads to a strong
increase of the EPC at {\bf K} as compared to the LDA and GGA functionals.
However, comparing to the GW value, the EPC is overestimated by 30\%
and the corresponding frequency for the {\bf K}-A$'_1$ mode at
1172 cm$^{-1}$ falls well below the degenerate {\bf K}-mode which
is around 1200 cm$^{-1}$ in the experiment \cite{maultzschPRL04,mohr07} 
(Fig.~\ref{fig1}) and at 1228 cm$^{-1}$ in our phonon calculation with B3LYP.
We have checked that tuning the percentage of HF-exchange in the hybrid 
functional allows to match the EPC value of the GW approach (in which case,
the {\bf K}-A$'_1$ mode remains the highest mode. This may be 
a good way to calculate the full phonon dispersion of graphite/graphene
within DFT, yet with an accuracy close to the one of the GW approach.

Concluding, GW is a general approach to compute accurate electron-phonon
coupling where DFT functionals fail.
Such a failure in graphite/graphene is due to the interplay between
the two-dimensional Dirac-like band structure and the
long-range character of the Coulomb interaction~\cite{basko08}.
However, GW can be also used in cases (in which the EPC is badly described by
DFT) where the electron-correlation is short ranged~\cite{zhang08}.

Calculations were done at IDRIS (081202, 081827).
C.A. and L.W. acknowledge French ANR PJC05\_6741.
We thank D.M. Basko, A. Rubio, J. Schamps, and C. Brouder for
discussions and A. Marini for the code {\tt Yambo}.


\begin{thebibliography}{99}

\bibitem{DFPT0}
S. Baroni, S. de~Gironcoli, A. Dal~Corso, and P. Giannozzi,
Rev. Mod. Phys. {\bf 73}, 515 (2001).

\bibitem{note_defs}
LDA, GGA and B3LYP refer, respectively, to
D.M. Ceperley and B.J. Alder, Phys. Rev. Lett. {\bf 45}, 566 (1980);
J.P. Perdew, K. Burke, and M. Ernzerhof,
Phys. Rev. Lett. {\bf 77}, 3865 (1996);
A.D. Becke, J. Chem. Phys. {\bf 98}, 5648 (1993).

\bibitem{GWreview}
F. Aryasetiawan and D. Gunnarsson, Rep. Progr. Phys. {\bf 61}, 237 (1998);
W.G. Aulbur, L. J\"onsson, and J.W. Wilkins, Solid State Phys. {\bf 54}, 1 (2000);
G. Onida, L. Reining, and A. Rubio, Rev. Mod. Phys. {\bf 74}, 601 (2002).

\bibitem{reich04}
S. Reich and C. Thomsen,
Phil. Trans. R. Soc. London A {\bf 362}, 2271 (2004).

\bibitem{ferrari06}
%A.C. Ferrari {\it et al.},
A.C. Ferrari, J.C. Meyer, V. Scardaci, C. Casiraghi, M. Lazzeri, F. Mauri, S. Piscanec, D. Jiang, K.S. Novoselov, and A.K. Geim, 
Phys. Rev. Lett. {\bf 97}, 187401 (2006);
%D. Graf {\it et al.} 
D. Graf, F. Molitor, K. Ensslin, C. Stampfer, A. Jungen, C. Hierold, and L. Wirtz,
Nano Lett. {\bf 7}, 238 (2007).

\bibitem{yao00}
Z. Yao, C. L. Kane, and C. Dekker,
Phys. Rev. Lett. {\bf 84}, 2941 (2000).

\bibitem{piscanec04}
S. Piscanec, M. Lazzeri, F. Mauri, A.C. Ferrari, and J. Robertson,
%S. Piscanec {\it et al.},
Phys. Rev. Lett. {\bf 93}, 185503 (2004).

\bibitem{maultzschPRL04}
J. Maultzsch, S. Reich, C. Thomsen, H. Requardt, and P. Ordej\'on,
%J. Maultzsch {\it et al.},
Phys. Rev. Lett. {\bf 92}, 075501 (2004).

\bibitem{mohr07}
% M. Mohr {\it et al.}
M. Mohr, J. Maultzsch, E. Dobard{\v z}i\'c,  I. Milo{\v s}evi\'c, 
M. Damnjanovi\'c, A. Bosak, M. Krisch, and C. Thomsen,
Phys. Rev. B {\bf 76}, 035439 (2007).

\bibitem{pocsik98}
I. P\'ocsik, M. Hundhausen, M. Ko\'os, and L. Ley,
J. Non-Cryst. Solids {\bf 227}, 1083 (1998).

\bibitem{tan02}
P. Tan, L. An, L. Liu, Z. Guo, R. Czerw, D.L. Carrol, P.M. Ajayan, N. Zhang, and H. Guo,
%P. Tan {\it et al.}
Phys. Rev. B {\bf 66}, 245410 (2002).

\bibitem{maultzschPRB04}
J. Maultzsch, S. Reich, and C. Thomsen,
Phys. Rev. B {\bf 70}, 155403 (2004).

\bibitem{thomsen00}
C. Thomsen and S. Reich,
Phys. Rev. Lett. {\bf 85}, 5214 (2000).

\bibitem{basko08}
D. M. Basko and I. L. Aleiner, Phys. Rev. B {\bf 77}, 041409(R) (2008).

\bibitem{DFPT1}
Calculations were done as in Ref.~\onlinecite{DFPT0}.
Technical details and thus phonon dispersions
are the same as in Ref.~\onlinecite{piscanec04}.

\bibitem{PBE_LDA}
LDA and GGA calculations were done with the code {\tt PWSCF}
(S. Baroni {\it et al.}, http://www.quantum-espresso.org),
with pseudopotentials of the type
N. Troullier and J. L. Martins, Phys. Rev. B 43, 1993 (1991).

\bibitem{zhou05}
%S.Y. Zhou {\it et al.}
S.Y. Zhou, G.H. Gweon, C.D. Spataru, J. Graf, D.H. Lee, S.G. Louie, and A. Lanzara
Phys. Rev. B {\bf 71}, 161403(R) (2005);
S.G. Louie, in Topics in Computational Materials Science,
C.Y. Fong editor (World Scientific, Singapore, 1997), p.96.

\bibitem{gruneis08}
%A. Gr\"uneis {\it et al.}
A. Gr\"uneis, C. Attaccalite, T. Pichler, V. Zabolotnyy, H. Shiozawa, S. L. Molodtsov, D. Inosov, A. Koitzsch, M. Knupfer, J. Schiessling, R. Follath, R. Weber, P. Rudolf, L. Wirtz, and A. Rubio
Phys. Rev. Lett. {\bf 100}, 037601 (2008).

\bibitem{GW}
GW calculations were done with the code {\tt Yambo}
(A. Marini et al., the {\tt Yambo} project, http://www.yambo-code.org/),
within the non-self consistent G$_0$W$_0$ approximation,
starting from DFT-LDA wave-functions and using a plasmon-pole model for
the screening, following 
M.S. Hybertsen and S.G. Louie, Phys. Rev. B {\bf 34}, 5390 (1986).

\bibitem{CRYSTAL}
HF and B3LYP calculations were done with the code {\tt CRYSTAL}
(V.R. Saunders et al. 
%, R. Dovesi, C. Roetti, R. Orlando, C.M. Zicovich-Wilson,
%N.M. Harrison, K. Doll, B. Civalleri, I.J. Bush, Ph. D’Arco, M. Llunell
{\tt CRYSTAL03} User’s Manual, University of Torino, Torino, 2003),
using the TZ basis by Dunning (without the diffuse
P-function). 

\bibitem{comp_det}
For graphite we used the experimental lattice parameters
($a$=2.46~\AA, $c$=6.708~\AA). For graphene we used $a$=2.46~\AA~
and a vacuum layer of 20~a.u.. EPCs were calculated on a structure
distorted by $d=$0.01 a.u.
For graphene, the electronic integration on the 1$\times$1 cell was
done with a 18$\times$18$\times$1 grid for LDA/GGA,
36$\times$36$\times$1 for GW, 66$\times$66$\times$1 for B3LYP/HF.
For graphite it was 18$\times$18$\times$6.
For the $\sqrt{3}\times\sqrt{3}$ cell we used the nearest equivalent k-grid.
Plane-waves are expanded up to 60 Ry cut-off.
We used a Fermi-Dirac smearing with 0.002 Ry width for
B3LYP/HF/GW and a Gaussian smearing with 0.02 Ry width for LDA/GGA.

\bibitem{slonczewski58}
J.C. Slonczewski and P.R. Weiss Phys. Rev. {\bf 109}, 272 (1958).
See also Suppl. information to
S. Pisana, M. Lazzeri, C. Casiraghi, K.S. Novoselov, A.K. Geim, A.C. Ferrari, and F. Mauri,
%S. Pisana {\it et al.}, 
Nature Materials {\bf 6}, 198 (2007).

\bibitem{note01}
In graphite, at the high-symmetry {\bf H} point 
the four $\pi$ bands are degenerate two-by-two,
$\Delta\epsilon_0$ being the energy difference.
By displacing the atoms according to the ${\bm \Gamma}$
E$_{2g}$ phonon, these bands remain degenerate and the energy
difference is increased by $\Delta\epsilon$.
In analogy to Eqs.~\ref{eqEPC1}-~\ref{eqEPC2} we define
$\overline{\langle D^2_{\bm \Gamma}\rangle}_{\rm F} =
(\Delta\epsilon^2-\Delta\epsilon_0^2)/(16 d^2)$.
By displacing the atoms according to the {\bf K}
A$'_1$ phonon, the four bands are no longer degenerate,
being $\pi^*$ ($\pi$) the two bands which are
up(down)-shifted and $\Delta\epsilon=\epsilon_{\pi^*}-\epsilon_\pi$.
We define
$\overline{\langle D^2_{\bf K}\rangle}_{\rm F} =
(\overline{\Delta\epsilon^2}-\Delta\epsilon_0^2)/(8 d^2)$,
where $\overline{\Delta\epsilon^2}$ 
indicates the average
between the four possible $\pi^*-\pi$ couples.

\bibitem{cumber}
In principle,
a direct calculation of $\omega_{\bf K}^{GW}$ (and thus of $B^{GW}_{\bf K}$)
could be obtained, e.g., by finite differences
from a prohibitively expensive GW total energy calculation.

\bibitem{zhang08}
P. Zhang, S.G. Louie and M.L. Cohen, Phys. Rev. Lett. {\bf 98}, 067005 (2007).


\end{thebibliography}
\end{document}